\documentclass[conference]{IEEEtran}
\IEEEoverridecommandlockouts

\usepackage{cite}
\usepackage{amsmath,amssymb,amsfonts}
\usepackage{color,soul}
\usepackage{algorithmic}
\usepackage{graphicx}
\usepackage{textcomp}
\usepackage{xcolor}
\usepackage{float}
\usepackage{todonotes}
\usepackage{url}
\usepackage{listings}
\def\BibTeX{{\rm B\kern-.05em{\sc i\kern-.025em b}\kern-.08em
    T\kern-.1667em\lower.7ex\hbox{E}\kern-.125emX}}
\begin{document}

\title{Streamlined Airborne Software Development for Large UAVs: From Unified Data Collection to Automated Code Generation\\

}
\author{
\centering
\IEEEauthorblockN{1\textsuperscript{st} Viktor Sinitsyn}
\IEEEauthorblockA{\textit{Institute of Flight System Dynamics} \\
\textit{Technical University of Munich}\\
Munich, Germany \\
viktor.sinitsyn@tum.de}
\and

\IEEEauthorblockN{2\textsuperscript{nd} Nils Schlautmann}
\IEEEauthorblockA{\textit{Institute of Flight System Dynamics} \\
\textit{Technical University of Munich}\\
Munich, Germany \\
nils.schlautmann@tum.de}
\\
\IEEEauthorblockN{4\textsuperscript{th} Prof. Dr.-Ing Florian Holzapfel}
\IEEEauthorblockA{\textit{Institute of Flight System Dynamics} \\
\textit{Technical University of Munich}\\
Munich, Germany \\
florian.holzapfel@tum.de}

\and

\IEEEauthorblockN{3\textsuperscript{rd} Florian Schwaiger}
\IEEEauthorblockA{\textit{Institute of Flight System Dynamics} \\
\textit{Technical University of Munich}\\
Munich, Germany \\
f.schwaiger@tum.de}
}

\maketitle

\begin{abstract}
The aerospace industry has experienced significant transformations over the last decade, driven by technological advancements and innovative solutions in goods and personal transportation. This evolution has spurred the emergence of numerous start-ups that now face challenges traditionally encountered by established aerospace companies. Among these challenges is the efficient processing of digital intra-device communication interfaces for onboard equipment—a critical component for ensuring seamless system integration and functionality. Addressing this challenge requires solutions that emphasize clear and consistent interface descriptions, automation of processes, and reduced labor-intensive efforts.

This paper presents a novel process and toolchain designed to streamline the development of digital interfaces and onboard software, which our team has successfully applied in several completed projects. The proposed approach focuses on automation and flexibility while maintaining compliance with design assurance requirements. 
\end{abstract}

\section{Introduction}

\subsection{Context, Challenges, and Goals} \label{LTF-UL}
Our team (Institute of Flight System Dynamics, Technical University of Munich) operates within the academic sector and is relatively small (50-60 persons in total with about 20 involved in a project). However, we collaborate extensively with industrial partners on numerous unmanned aerial vehicles (UAV) projects, covering a wide range of aircraft types — fixed-wing aircraft, helicopters, electric vertical take-off and landing aircraft (eVTOL), and airships — with maximum take-off weights up to 2.7 tons. A key distinction of our work is that most research outcomes are implemented in real, flying aircraft in house. Beyond developing novel designs, navigation techniques, and control laws, our team’s responsibilities also include avionics interface design, system integration, and verification tasks.

While most of our projects involve UAVs, and thus do not formally require certification or full design assurance by regulation, the size and concepts of operations (ConOps) of these aircraft impose significant requirements, including:
\begin{itemize}
	\item Redundant and sophisticated avionics architectures,
	\item Safety assessments and design assurance activities, driven internally rather than by external regulations.
\end{itemize}

As a result, we face challenges typically associated with a larger aerospace companies:
\begin{itemize}
    \item Complexity: Large UAVs exhibit system complexity comparable to that of CS-23 Level 4 aircraft.
    \item Safety requirements: High internal standards must be met to ensure operational safety.
    \item Process transparency: Development processes must be clear and auditable.
    \item Verifiability of design artifacts: Artifacts must be organized to allow potential future certification efforts.
\end{itemize}
These challenges are accompanied by resource limitations as our team's size and funding are constrained compared to major aerospace manufacturers. Given these challenges, the goals for our development process are:
\begin{itemize}
	\item Efficiency: Achieved by maximizing automation of development activities and reusability of artifacts across multiple projects.
	\item Transparency and verifiability: Design a workflow that, while not immediately compliant with standards such as ARP4754B or DO-178C, can be incrementally upgraded to achieve full Design Assurance compliance without major refactoring.
\end{itemize}

\clearpage
\begin{figure*}[ht]
    \centering
    \includegraphics[width=0.99\textwidth]{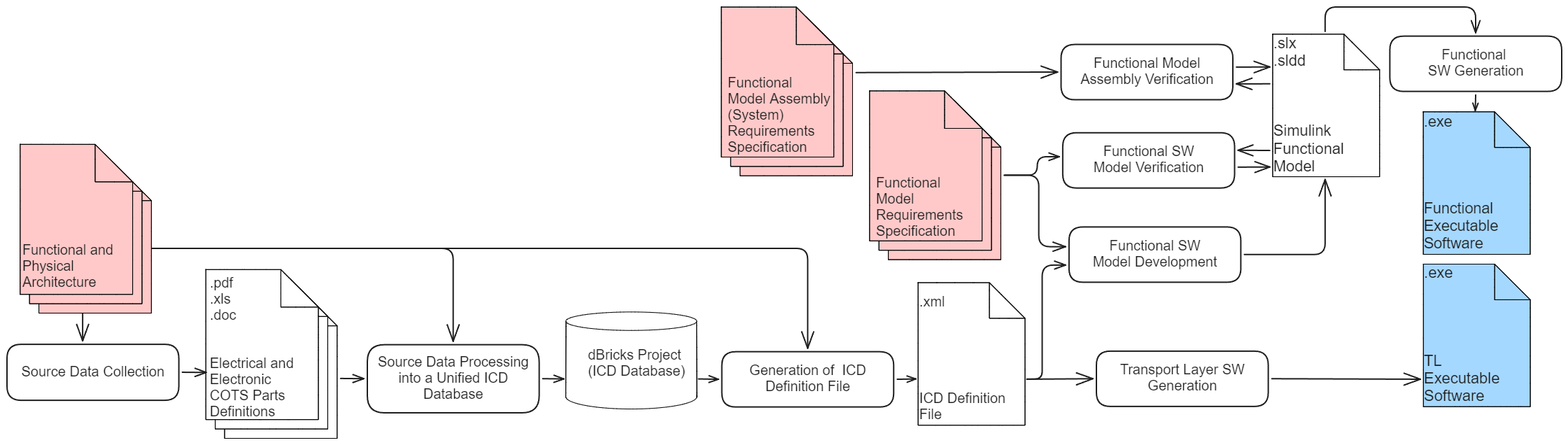}
    \caption {Process Overview }
    \label{fig:Process_overview}
\end{figure*}

\subsection{State of the Art}
The aerospace industry is steadily moving toward more automated, and model-driven processes for avionics and airborne software development. As systems increase in complexity, both academic and industrial communities have proposed workflows and toolchains aiming to improve traceability, reduce manual effort, and facilitate design assurance. This section highlights selected relevant works, grouped by key technical domains.
\begin{enumerate}
    \item Automated Handling of Interface Control Document (ICD) Data: Precise and traceable management of interface definitions is essential for modern avionics systems. Several approaches focus on creating structured, machine-readable ICDs and aligning them with architecture models.
    \begin{itemize}
        \item The eSAM method, introduced in \cite{b1} and extended in \cite{b2}, provides a structured approach for modeling data-message flows in IMA architectures, enabling transformation between logical, transport, and physical layers.
        \item In \cite{b26} an architecture-driven and web-enabled ICD configuration framework is presented. The approach includes domain-specific query languages and a flexible tool integration model for platform-level modeling and export.
        \item A similar layered ICD database concept is described in \cite{b41}, where machine-readable ICDs form the basis for automated code generation and platform configuration.
    \end{itemize}
    These efforts reflect a growing recognition of ICD data as a central artifact in both model-based development and downstream automation.
    \item Automated Code Generation in Safety-Critical Software: Model-based design and code generation have become increasingly viable for airborne software, especially when integrated with tools supporting verification and traceability.
    \begin{itemize}
        \item In \cite{b49} and \cite{b51}, a development team proposes a toolchain based on Simulink, Embedded Coder, and Polarion. The solution includes in-house tools such as SimPol and mrails, which manage traceability and build automation to meet DO-178C and DO-331 requirements.
        \item The process detailed in \cite{b33} describes a lean, certifiable model-based software workflow that minimizes overhead while remaining aligned with formal standards. The workflow supports agile iteration cycles and formal artifact generation.
    \end{itemize}
    These studies collectively demonstrate that automated generation of safety-critical software is achievable, even in regulated domains, when supported by traceable models and well-defined toolchains.
    \item Toolchain Integration via Domain-Specific Tools: Beyond single-tool workflows, several initiatives aim to integrate domain-specific tools into cohesive development environments.
        \begin{itemize}
        \item As mentioned above, the team behind \cite{b26} and \cite{b29} proposes using a query-driven, model-integrated environment to bridge architectural and platform-specific data. Their web-based interface allows for real-time artifact management and cross-tool synchronization.
        \item The AvioNET framework \cite{b38} focuses on building toolchains through data transformation pipelines connecting various domain-specific languages (DSLs), supporting highly automated avionics development.
        \item In \cite{b40} formal methods and model checking are introduced into the avionics software lifecycle, demonstrating how safety-critical assurance processes can be integrated into development flows.
        \item The platform-level design methodology presented in \cite{b48} explores how multi-tool automation and structured model reuse can reduce effort and improve consistency across complex avionics programs.
    \end{itemize}

\end{enumerate}
In addition to these technical implementations, several broader studies provide context and motivation for toolchain evolution:
\begin{itemize}
    \item Overviews such as \cite{b7} summarize development challenges in modern avionics, especially with respect to aligning innovation with certification and managing distributed development.
    \item A stakeholder-centered analysis in \cite{b52} identifies tooling gaps and workflow pain points in Integrated modular avionics (IMA) development environments.
    \item A system-level perspective on agile-compatible architecture development based on ARP4754 is presented in \cite{b53}, offering insight into how early design phases can feed into model-based workflows.
\end{itemize}

Taken together, these works provide a rich foundation for understanding the state of digital avionics development. However, from the perspective of lightweight and scalable toolchain design for experimental UAVs, several practical gaps remain:

Many proposals are conceptual in nature, offering valuable modeling formalisms and platform definitions but lacking reproducible real-life implementations. Others rely on proprietary tools that, while powerful, are inaccessible for academic teams. Much of the published work is also tightly coupled with IMA-based system architectures, whereas our focus lies in federated platforms typically used in cost-sensitive UAV projects.

Perhaps most notably, despite significant progress across individual domains—ICD modeling, code generation, tool integration—there is a lack of end-to-end, lightweight toolchains that combine these elements into a coherent and practical development process.

Building on the goals and gaps identified in the state of the art, this work presents a modular, automation-oriented process and toolchain tailored for airborne software development in large UAV platforms. The process is grounded on three guiding principles: automation of all feasible process steps using domain-specific tools; support for machine-readable intra-tool data exchange to ensure consistency and traceability; and explicit definition of intermediate artifacts to enhance transparency and flexibility. By adhering to these principles, we achieve a workflow that is highly automated, modular and scalable across varying project sizes and criticality levels, and structured in a way that can be extended toward formal design assurance with minimal refactoring.

\subsection{Outline}
The remainder of this paper is organized as follows. Section II provides an overview of the proposed development process, highlighting its main components, the flow of artifacts, and their interdependencies. Section III offers a detailed description of each process step, including the roles of domain-specific tools, the structure of intermediate artifacts, and the mechanisms used to ensure automation and traceability. Section IV presents implementation examples drawn from actual UAV projects. Finally, Section V summarizes the results and outlines directions for future development.

\section{Toolchain and process overview} \label{section:certification_basis}

The proposed process overview is shown in Figure~\ref{fig:Process_overview}.
The final goal of the proposed development process is to produce two major categories of software artifacts:
\begin{itemize}
    \item Functional Software -  e.g. flight control applications that compute actuator commands based on operator inputs and sensor data.
    \item Transport Layer (TL) Software - middleware that decodes incoming data from digital buses (e.g. sensors signals) and encodes outgoing data (e.g. control commands) into protocol-specific messages.
\end{itemize}
These two artifacts are highlighted in blue on the right-hand side of the process overview diagram.

On the input side, the process begins with three types of requirement specifications:
\begin{itemize}
    \item Functional and Physical Architecture Documents, describing the aircraft functions grouping, relations and mapping to devices.
    \item Functional Model Requirements, applicable to a specific functional application (e.g. flight control primary application).
    \item Functional Model Assembly Requirements, applicable to a system or complete aircraft.
\end{itemize}
Of these, the Architecture and Single Functional Model Requirements are directly used during development, while the Functional Model Assembly Requirements primarily support model verification prior to integration. These input artifacts are marked in pink on the left-hand side of Figure~\ref{fig:Process_overview}.
In the diagram, processes are depicted as round-edged rectangles, and artifacts are represented with corresponding icons.
\subsection{Source Data Collection}
The process begins by gathering and organizing the source data necessary for interface development.
\subsection{Source Data Processing into a Unified ICD Database}
Collected data is processed into a unified database that consolidates all interface information. 

\subsection{Generation of ICD Definition File in XML Format}
The unified ICD database serves as the basis for generating machine-readable and human-verifiable XML-based intermediate ICD files for a specific project-tailored devices.

\subsection{Functional Software Model Development}
Application-layer functional software is modeled using MathWorks Simulink.
Interfaces between functional and transport layers are explicitly defined through the generated XML ICD definitions, ensuring seamless integration and simplifying subsequent steps like code generation and testing.
\subsection{Functional SW Model Verification}
Developed Simulink models undergo verification activities, including:
\begin{itemize}
    \item Standalone testing of individual functional models
    \item Integration testing based on the Functional Model Assembly Requirements
\end{itemize}
Verification activities ensure early detection of design flaws before software implementation.

\subsection{Functional SW Generation}
Following model verification, code generation from Simulink models is automated using the MathWorks Embedded Coder toolset.
Depending on the criticality level of the software, additional automated verification steps may be incorporated into the workflow. 
It is worth highlighting that as code generation process is fully automated change management process is reduced to automatic regeneration of the source code and all relevant artifacts whenever a new model revision becomes available. 

\subsection{Transport Layer SW Generation}
The transport layer software bridges functional application input/output variables with transport-specific message formats. Using the same XML-based ICDs as the source information, a universal code generator produces transport-layer software based on user-defined templates. The process can be fully automated based on the project needs resulting in a reduced changed management process similar to one explained above.
\section{Detailed toolchain and processes description}

\subsection{Source Data Collection}
We define source data collection as a distinct process step for the following reasons:
\begin{enumerate}
    \item Incomplete and error-prone vendor data: In the UAV domain, vendor-supplied documentation is often incomplete, ambiguous, or inaccurate. Clarifications must typically be obtained through direct communication with vendors. Without a structured collection process, important details were historically lost over time, causing issues when reusing data in subsequent projects.
    \item Baseline and change management necessity: Maintaining accurate, versioned source data with clear baseline control is critical for mid- and long-term project continuity.
    \item Single Source of Truth: Establishing a centralized, structured database ensures all team members access consistent and verified device interface information.
\end{enumerate}
\begin{figure}[ht]
    \centering
    \includegraphics[width=0.49\textwidth]{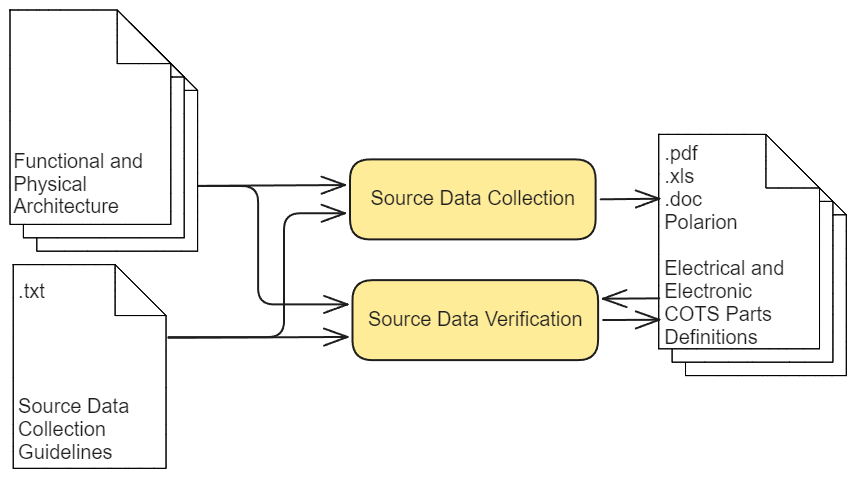}
    \caption {Source Data Collection}
    \label{fig:Source_Data_Collection}
\end{figure}
An overview of the source data handling process is shown in Figure~\ref{fig:Source_Data_Collection}.
Here and in following sections process steps are depicted as rounded rectangles. \textbf{Manual steps are colored yellow, automatic - green}.
The process consists of two main steps:
\begin{enumerate}
    \item Source Data Collection: Gathering all relevant vendor documentation and supplementary information,
    \item Source Data Verification: Optional step of reviewing and validating data completeness and integrity.
\end{enumerate}

\begin{figure}[ht]
    \centering
    \includegraphics[width=0.49\textwidth]{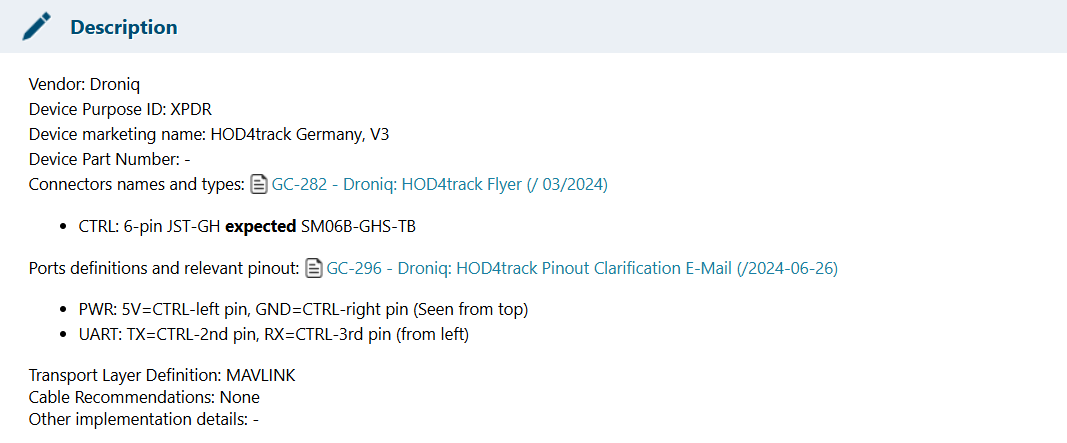}
    \caption {Linked Dataset Description Screenshot}
    \label{fig:Linked_Dataset_Description_Screenshot}
\end{figure}

For each device the following minimum information set is collected and maintained:
\begin{enumerate}
    \item Source files received from vendors,
    \item Source files metadata including following data when applicable and available:
    \begin{itemize}
        \item Name,
        \item Part number and revision,
        \item Release date and institution,
        \item File delivery source,
        \item etc.
    \end{itemize}
    \item Explicit References for Key information:
     \begin{itemize}
        \item Physical connectors names and part numbers,
        \item Physical ports pinouts,
        \item Transport layer details.
    \end{itemize}   
\end{enumerate}
The results of the data collection process are stored as "linked datasets" associated with "Part" configuration items in the Configuration Management (CM) database. We use Siemens Polarion \cite{o1} as the backbone for our configuration management, enabling:
\begin{itemize}
    \item Traceable baselines for all devices,
    \item Change tracking across project lifecycle,
    \item Integration with other development artifacts.
\end{itemize}
Linked dataset description example is shown in Figure ~\ref{fig:Linked_Dataset_Description_Screenshot}

\subsection{Source Data Processing into a Unified ICD Database}
The next step of the process takes the source data collected in the previous stage and combines it with system architecture information to create a unified ICD database.

\begin{figure}[ht]
    \centering
    \includegraphics[width=0.49\textwidth]{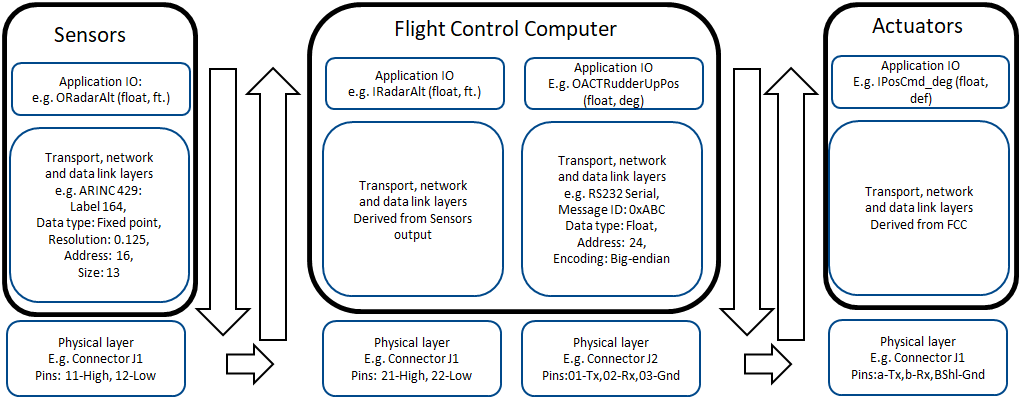}
    \caption {dBricks Data Model}
    \label{fig:dBrDataModel}
\end{figure}

We use a domain-specific tool called dBricks, a client-server relational database system with a web-based front-end.
dBricks is specifically designed to manage ICD data across physical, transport, and logical layers. It implements a normalized relational data model, enabling consistent definition of device templates, project-specific device instances, and their interconnections. The tool supports continuous integrity checking and integration with external development environments via a Representational State Transfer Application Programming Interface (REST API). A detailed description of the tool’s data model is available at \cite{o2}. Use case example is illustrated in \cite{b41}.
An overview of the dBricks data model relevant to our development process is shown in Figure \ref{fig:dBrDataModel}.

For the purpose of our workflow description, the database organizes ICD data into three interconnected layers:
\begin{itemize}
    \item Physical Level: Includes information on connectors, pinouts, and inter-device wiring.
    \item Transport Level: Defines how logical parameters are encoded into messages transmitted over serial protocols.
    \item Logical Level: Describes the structure, composition, and data types of inputs and outputs for functional applications.
\end{itemize}
The dBricks tool applies a data normalization principle of reducing data redundancy to maximize consistency, modularity, and reuse across projects. At the core of the dBricks data model lies a strict separation between device templates and device instances as project parts. Templates define generic, reusable device descriptions independent of any specific project. They contain all necessary information about a device's logical, transport, and physical interfaces.
In contrast, device instances are created by referencing templates and adapting them to a specific project’s architecture, without duplicating data.
Thus, the development workflow in dBricks follows these steps:
\begin{enumerate}
    \item Commercial of-the-Shelf (COTS) Device Template Creation: For each COTS device, a device template is created based on the collected source documentation.
    \item Project-Specific Device Template Derivation: Project-specific device templates are created by adapting connected COTS devices configuration to reflect the specific system architecture and project requirements.
    \item System Architecture Refinement (parallel activity): When derived interface requirements (e.g., project-specific cross-talk message structures) emerge, they are documented in the updated System Architecture Documents. For simplicity, these steps are not explicitly shown in the process diagram.
    \item Project Assembly: All device descriptions and their physical and logical interconnections are gathered into a Project entity within dBricks.
\end{enumerate}

\begin{figure}[ht]
    \centering
    \includegraphics[width=0.49\textwidth]{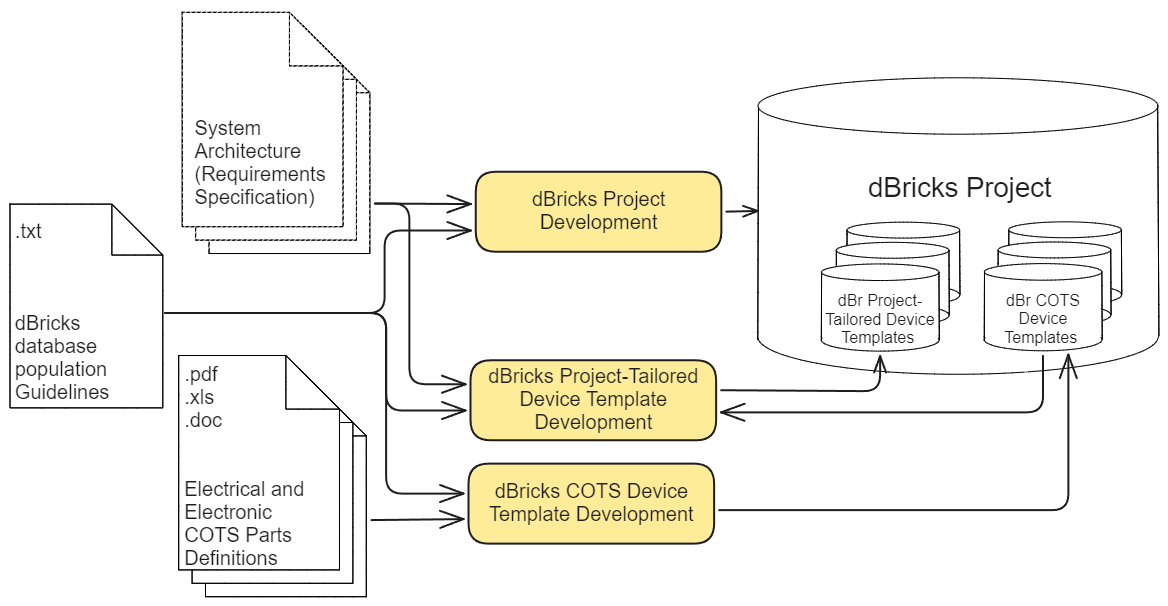}
    \caption {Source Data Processing into a Unified ICD Database}
    \label{fig:ICD Database Population}
\end{figure}

The overall process is illustrated in Figure~\ref{fig:ICD Database Population}.
Once the project structure is complete, the dBricks project can undergo optional verification and baselining.

\subsection{Generation of ICD Definition File in XML Format}
After developing the project-specific ICD database in dBricks, the next step is the generation of intermediate artifacts that are both machine-readable and human-verifiable.

Although dBricks supports several tool connectivity options, including direct API-based interaction, we selected the export of a user-configured XML-based ICD definition file for the following reasons:
\begin{itemize}
    \item Configuration Management: The XML file is a discrete, easily controlled configuration artifact, which can be independently baselined, versioned, and audited.
    \item Independent Verification: The XML artifact can be independently verified against the original source data (vendor documentation) and the defined system architecture.
    \item Toolchain decoupling: XML files decouple toolchain components, offering flexibility in tool selection and integration. For example, subcontractors can provide XML files in the prescribed format, or alternative tools can be seamlessly integrated into the workflow.
\end{itemize}

The XML export format is fully user-configurable through a set of SQL queries that access the dBricks database directly.  
Each SQL query set can be configured and baselined as a separate artifact.

The XML file developed for our projects contains the following main components:
\begin{itemize}
    \item Functional Input/Output Definitions: Logical signals and data structures used by functional software models.
    \item Physical Port Definitions: Connector and pinout information for physical interfacing between devices.
    \item Transport Layer Specifications: Mapping of logical I/O signals onto communication protocols, including message formats, data encoding (e.g., data types, endianness), and protocol metadata (e.g. message identifiers, baud rate etc.).
    \item Data Type Definitions: Definition of essential properties for atomic data types and structure for complex data types.
\end{itemize}

A simplified excerpt from an example FCCN ICD XML file is shown in Listing~\ref{lst:icdexample}.

\begin{lstlisting}[language=XML, caption={Excerpt from generated ICD XML artifact}, label={lst:icdexample}, breaklines=true, showstringspaces=false, basicstyle=\ttfamily\scriptsize]
<root>
 <Devices>
  <Device name="Flight Control Computer Normal" id="FCCN">
   <Functions>
    <Function name="in" layer="Development">
     <Parameters>
      <Parameter name="OCE_Cmds" direction="out" 
       data_type="DCE_Cmds"/>
     </Parameters>
    </Function>
   </Functions>
   <Ports>
    <Port name="DSGCAN01" bus_type="SGCAN"
          direction="duplex">
     <Contacts>
      <Contact wire="Hi" connector="ST12" contact="11"/>
      <Contact wire="Lo" connector="ST12" contact="02"/>
      <Contact wire="Sh" connector="ST12" contact="BShl"/>
     </Contacts>
    </Port>
   </Ports>
   <Port_Contents>
    <Port_Content name="DSGCAN_CAN1" direction="out">
     <Frames>
      <Frame name="F_ACTFLO_Cmd_Pos" size="83" 
        transmit_rate_ms="10" type="CAN_SF">
       <IDs>
        <Container name="ID" address="0" value="0x60A"/>
        <Container name="RTR" address="11" value="0"/>
       </IDs>
       <Payload>
        <Container name="Payload_Pos" address="51"
                   linked_parameter="in.OCE_Cmds.
                   ACTFXX.ACTFLO_Cmd_Pos_rad"/>
       </Payload>
      </Frame>
     </Frames>
    </Port_Content>
   </Port_Contents>
  </Device>
 </Devices>
 <DataTypes>
  <DataType name="DCE_Cmds" type="Complex" size="384">
   <Elements>
    <Element name="ESCXX" address="0"/>
    <Element name="ACTFXX" address="96"/>
    <Element name="ACTTXX" address="240"/>
    <Element name="ACTRX" address="336"/>
   </Elements>
  </DataType>
 </DataTypes>
</root>
\end{lstlisting}

\begin{figure}[ht]
    \centering
    \includegraphics[width=0.49\textwidth]{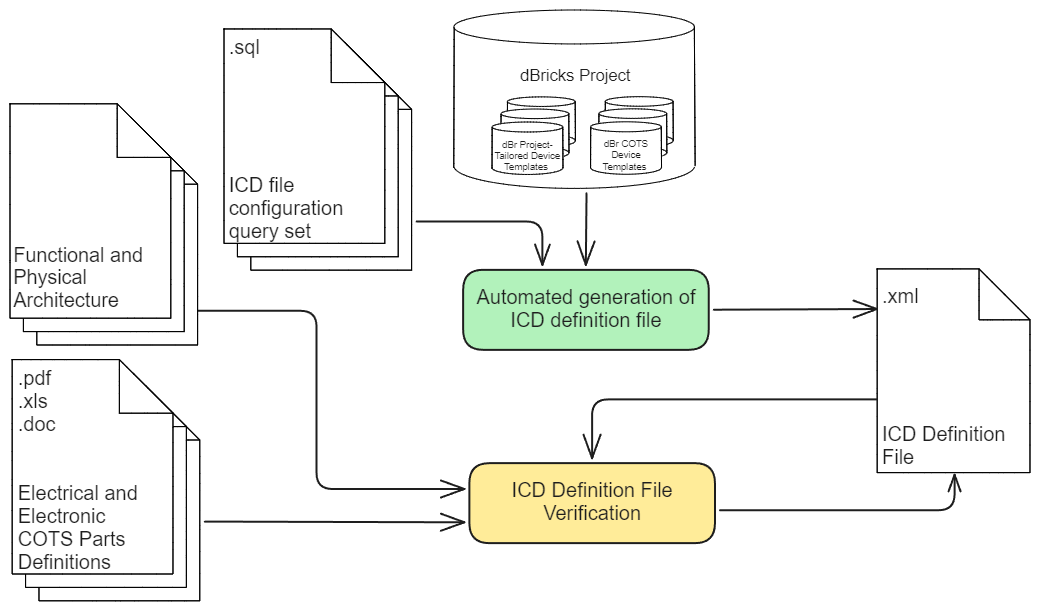}
    \caption {Generation of ICD Definition File in XML Format}
    \label{fig:xml_generation}
\end{figure}

It is important to note that the specific choice of XML format is not critical by itself. What matters is that the relevant development tools support standardized, machine-readable data exchange.  
In our process key compatibility points include:
\begin{itemize}
    \item dBricks: Supports exporting artifacts in XML, JSON, or XLS formats, configurable according to user needs.
    \item MathWorks Environment: Provides mature support for XML parsing and automated generation of Simulink model interfaces and data dictionaries.
    \item Transport Layer Code Generator: In-house development of a universal code generator consuming the XML ICD file to produce transport software code templates.
\end{itemize}
For safety-critical applications, additional formal verification steps are introduced:
\begin{itemize}
    \item The exported ICD XML file is manually verified against original COTS device documentation and the system architecture document.
    \item Due to the variability in vendor formats, full automation of verification is currently infeasible.
    \item To support manual verification, we are developing an XML Viewer Tool that renders the structure of the ICD file in a user-friendly, graphical interface and implements some additional productivity-related features to facilitate efficient review.
\end{itemize}
The detailed process steps are depicted in Figure ~\ref{fig:xml_generation}

\subsection{Functional SW Model Development}

\begin{figure}[ht]
    \centering
    \includegraphics[width=0.49\textwidth]{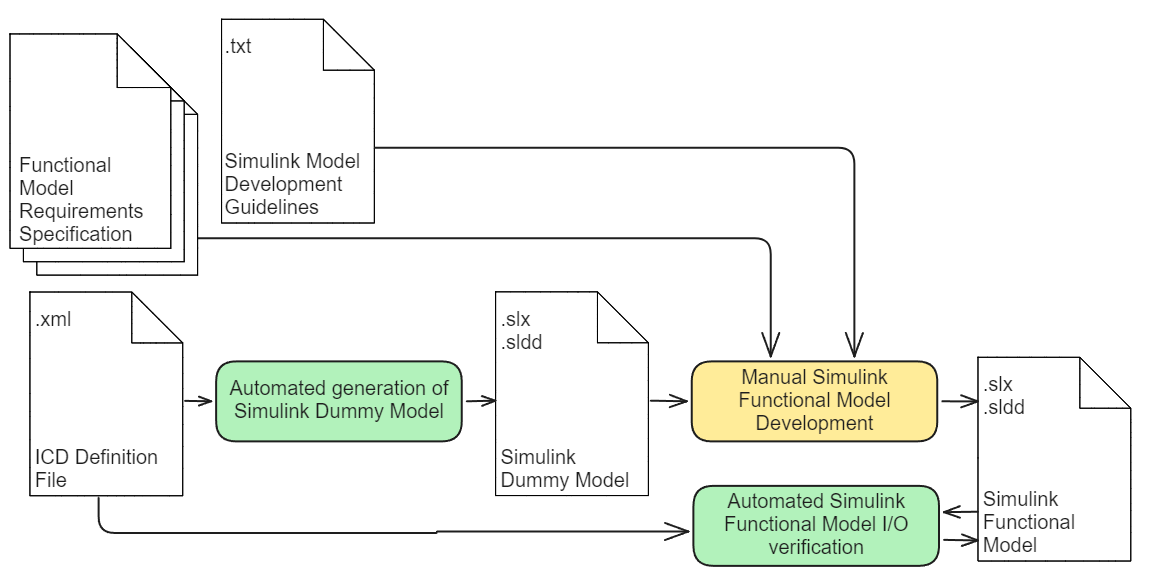}
    \caption {Functional SW Model Development}
    \label{fig:simulink-development-process}
\end{figure}

The process steps for functional software model development are illustrated in Figure~\ref{fig:simulink-development-process}.

The first step is to import the definitions of input/output (I/O) structures from the XML-based ICD file generated in the previous process stage.  
For each parameter, the following information is extracted:
\begin{itemize}
    \item Name,
    \item Direction (input or output),
    \item Data type for atomic parameters or structure definition for complex parameters (Simulink bus structure).
\end{itemize}

This import process results in the creation of a so-called "dummy model", consisting of a Simulink model file (.slx), and an associated Simulink data dictionary file (.sldd).

The dummy model contains only the I/O interface definitions and serves as the foundation for subsequent model development.

\begin{figure}[ht]
    \centering
    \includegraphics[width=0.49\textwidth]{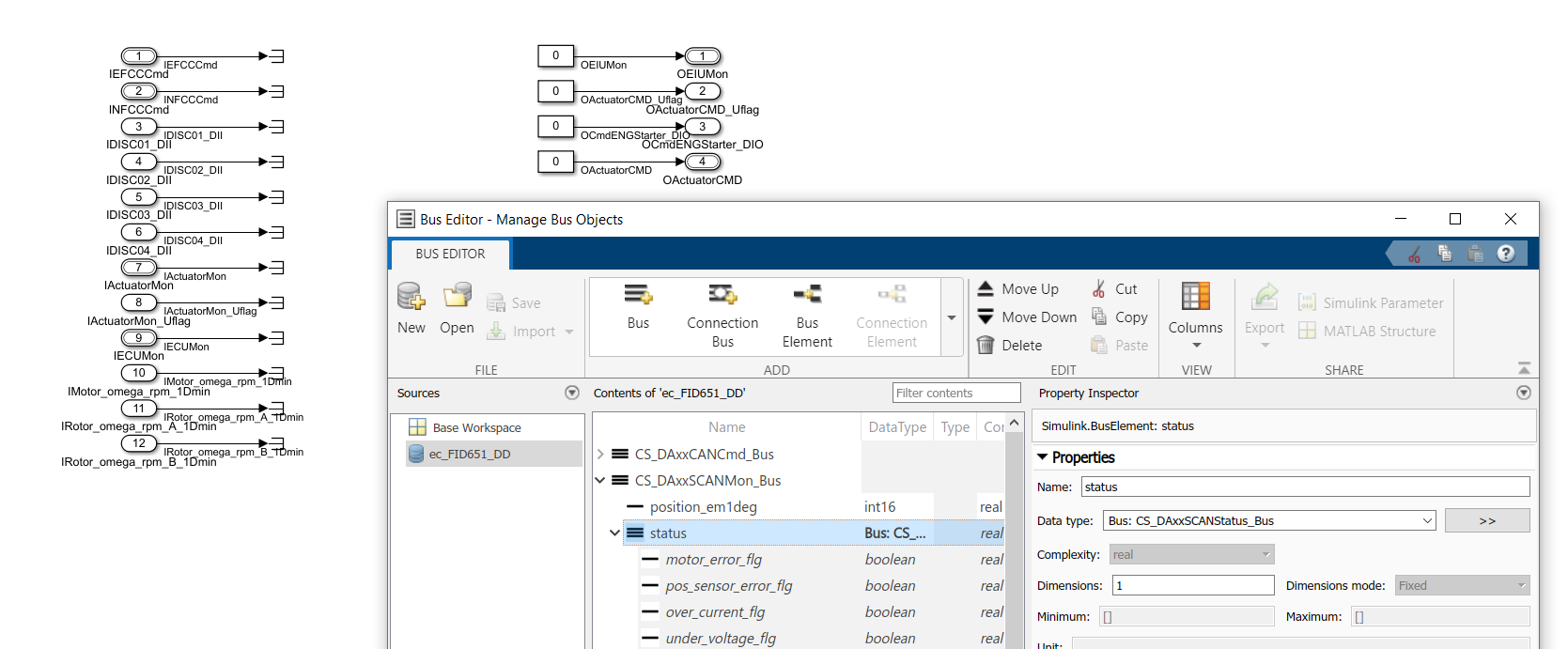}
    \caption {Functional Software Model Development}
    \label{fig:dummy_model_example}
\end{figure}

Example of a "dummy model" is shown in Figure~\ref{fig:dummy_model_example}.

Following the generation of the dummy model, the manual development of the actual functional model begins.  
The developer uses:
\begin{itemize}
    \item The dummy model for standardized I/O structure references,
    \item The Functional Model Requirements Specification as the source for functional behavior.
\end{itemize}

Any changes to the I/O structures must not be performed manually in the functional model as modifications outside the automated import process can lead to inconsistencies during system integration.

To enforce consistency, an automated I/O structure verification step is performed, checking the developed functional model against the original dummy model to ensure structural integrity.

\subsection{Functional SW Model Verification}

Once the functional model development is completed, the model undergoes verification activities.  
We currently apply two complementary verification techniques:
\begin{itemize}
    \item Manual testing,
    \item Automated testing.
\end{itemize}

Manual testing is performed based on predefined test procedures, with formal test reports generated as verification evidence.

Automated testing requires a more sophisticated setup, including the configuration of Simulink test harnesses and supporting toolchain pipelines.  
While the initial setup effort is higher, automated testing offers significant long-term efficiency gains, particularly for models with extended lifecycles or frequent modification cycles.

The trade-off between manual and automated verification depends primarily on two factors: the expected lifecycle duration of the model, and the criticality level of the modeled functionality. It is quite common that initial manual testing is transitioned to automatic testing together with project maturity growth.

In addition to verifying individual functional models, we also perform verification of models as part of larger model assemblies up to the complete closed Model-in-the-Loop testing of entire aircraft.  
The verification approach remains fundamentally similar; the primary difference lies in the level of the requirements specification documents used as source data.

\begin{figure}[ht]
    \centering
    \includegraphics[width=0.49\textwidth]{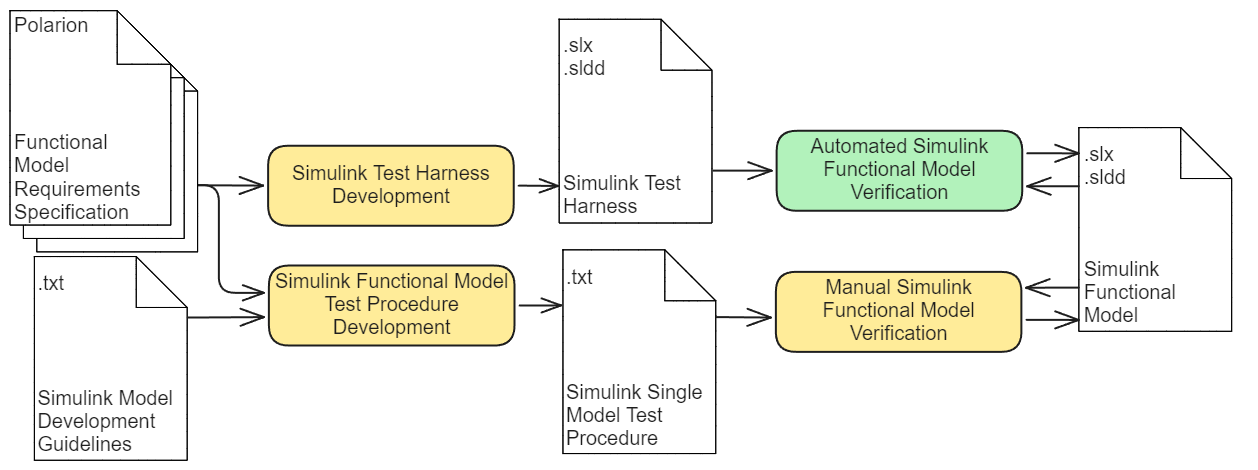}
    \caption {Single Functional SW Model Verification}
    \label{fig:single-model-verification}
\end{figure}

\begin{figure}[ht]
    \centering
    \includegraphics[width=0.49\textwidth]{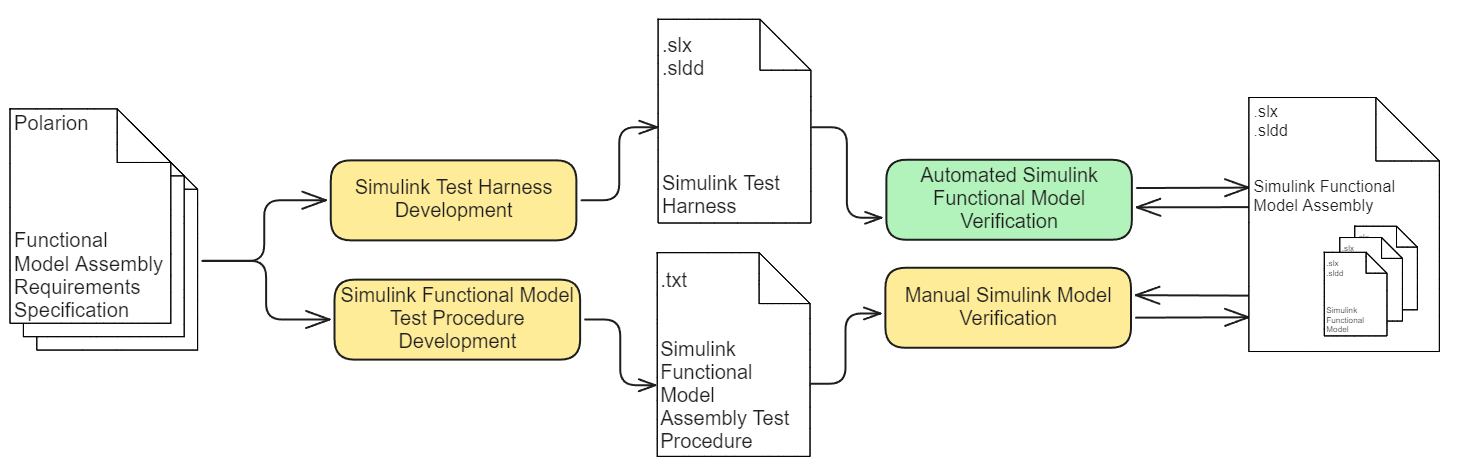}
    \caption {Functional Model Assembly Verification}
    \label{fig:model-assembly-verification}
\end{figure}

The process steps for single model verification and model assembly verification are depicted in Figures~\ref{fig:single-model-verification} and~\ref{fig:model-assembly-verification}, respectively.

\subsection{Functional SW Generation}

\begin{figure}[ht]
    \centering
    \includegraphics[width=0.49\textwidth]{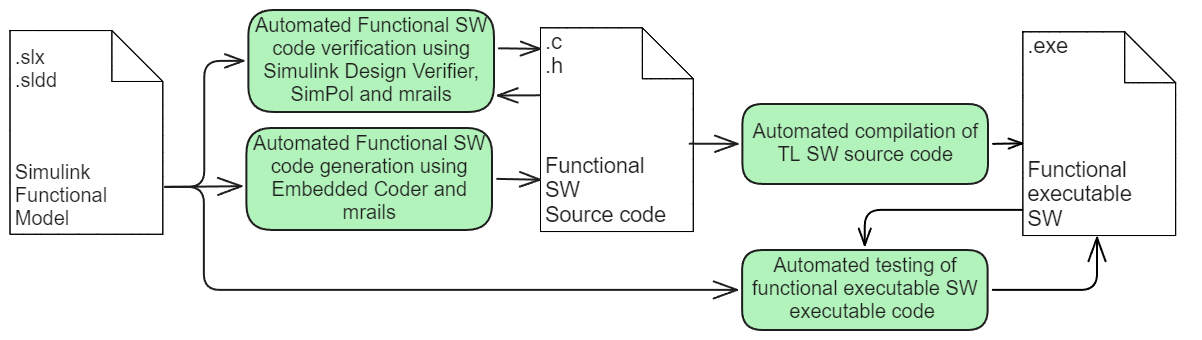}
    \caption {Functional SW Generation}
    \label{fig:functional-sw-development}
\end{figure}

The automated generation of functional SW is based on a model-driven development process centered around Simulink models. A simplified overview of the process steps is shown in Figure~\ref{fig:functional-sw-development}.

The process relies on the following key tools:
\begin{itemize}
    \item MathWorks Embedded Coder for code generation,
    \item Simulink Design Verifier and SimPol for model verification and compliance checks,
    \item Siemens Polarion together with SimPol for requirements traceability,
    \item mrails for process automation and build orchestration.
\end{itemize}

A detailed description of a complete, formally approvable safety-critical model-based development process is provided in \cite{b51}, \cite{b49}, \cite{b33} and is beyond the scope of this article.  
Here, we highlight that automated software generation and verification is feasible even for applications subject to design assurance, and that a simplified version of the process is highly effective for less critical projects.

The core input to the automated development process is a verified Simulink Functional Model, consisting of a Simulink model file (.slx) and associated Simulink data dictionary file (.sldd).

The following steps are performed during the automated functional software development:
\begin{itemize}
    \item Code generation is performed using Embedded Coder and the mrails tool, automatically producing source code files (.c, .h) from the Simulink model.
    \item Following code generation, the generated source code is verified against the Simulink model. In this process, the model serves as the authoritative low-level requirements (LLR).
    \item After verification, the source code is compiled into executable binaries using the appropriate platform-specific toolchain.
    \item Finally, automated tests are executed on the compiled software to validate its behavior against functional expectations. Test scenarios are derived from model-based test procedures and are typically integrated into continuous testing pipelines.
\end{itemize}

\subsection{Transport Layer SW Generation}

\begin{figure}[ht]
    \centering
    \includegraphics[width=0.49\textwidth]{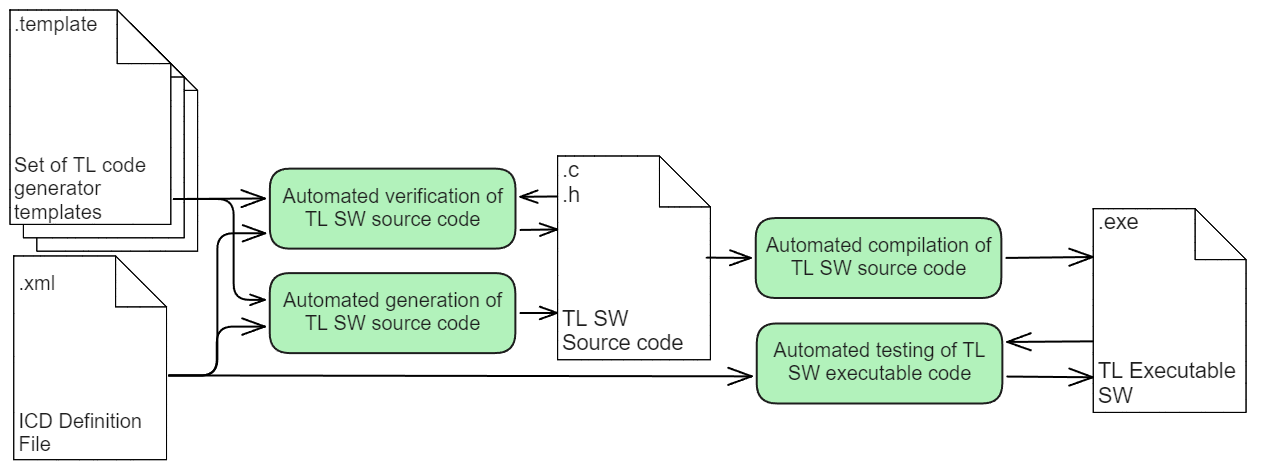}
    \caption {TL SW Generation}
    \label{fig:tl_sw_development}
\end{figure}

The final process in our toolchain is the automated generation of transport layer software.  
An overview of the process steps is shown in Figure~\ref{fig:tl_sw_development}.

The ICD definition files serve as a low-level requirements specification for the SW development process.  
They explicitly define how input data flows are mapped into input structures for the functional software and how output messages are encoded based on functional software outputs.

The core technology enabling this process is an automatic code generator developed in-house.  
The tool applies a generic ``text template rendering engine'' to the ICD data model, allowing for flexible and reusable code generation across different platforms.

The necessary data model is parsed from the XML-based ICD definition file, while the second critical component is a set of platform-specific templates.  
Since transport layer code is inherently platform-specific, a dedicated set of templates must be developed for each target platform.  
However, template development is a one-time activity for each platform type and can be reused across multiple projects.

For applications that do not require formal design assurance, this automated toolchain already delivers reliable results.  
Potential issues are typically detected during integration testing and corrected in the code generator templates, ensuring they do not reoccur in subsequent projects.

For safety-critical applications, however, formal verification and testing of the generated TL software are required.  
Manual verification of the generated code is labor-intensive and error-prone, motivating the development of automated verification and testing capabilities.

Both features are currently under active development:
\begin{itemize}
    \item Automated Verification is based on the idea of mapping generated code sections back to corresponding parts of the ICD definition file.  
    This requires additional markup within the templates to maintain traceability across the XML-to-source code transformation.
    
    \item Automated Testing is based on comparing the expected message bitstreams (defined by ICD specifications) with the bitstreams produced by the generated code functions.
\end{itemize}

If these developments prove successful, we plan to formalize the toolchain towards TQL-5 tool qualification, enabling the generated TL software to be eligible for integration into airborne systems at up to Design Assurance Level (DAL) C.

\section{Implementation Examples} 
The process described above was successfully implemented either fully or partially in several UAV and manned aircraft projects our team was involved in:
\subsection{EPUCOR}
EPUCOR is a coaxial rotorcraft CoAX 600 manufactured by edm aerotec GmbH \cite{o3} with maximum take-off weight (MTOW) of 600 kg. converted into an unmanned demonstrator platform with the aid of robotic actuator system installed in the pilot position. It serves as a reversible, full-scale testbed for validating flight control functions in real flight conditions. The project emphasizes rapid system integration, and experimental flexibility. Further details can be found in \cite{b54}
\begin{figure}[ht]
    \centering
    \includegraphics[width=0.49\textwidth]{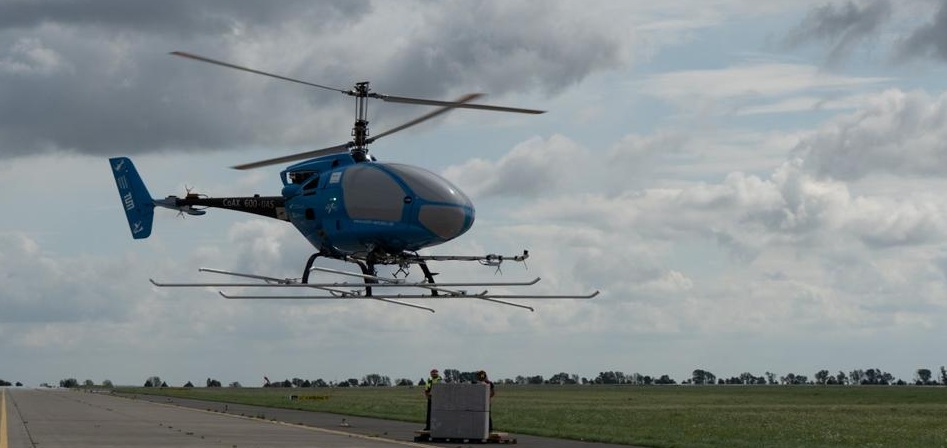}
    \caption {EPUCOR Helicopter}
    \label{fig:EPUCOR}
\end{figure}

\subsection{eRC's full scale demonstrators Echo and Romeo}
eRC \cite{o4} is developing manned lift-and-cruise aircraft with a MTOW of 2700 kg and either fully electrical or hybrid-electric propulsion. It targets the patient transportation niche currently occupied by conventional helicopters. The aircraft is a complex, safety-critical system developed with future certification in mind. The prototypes are currently undergoing initial flight tests.
\begin{figure}[ht]
    \centering
    \includegraphics[width=0.49\textwidth]{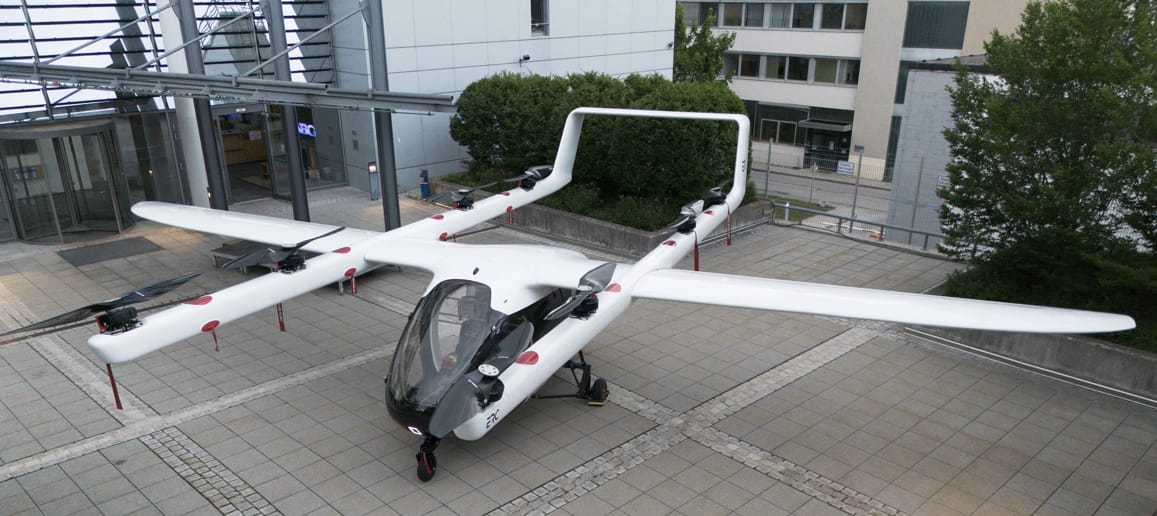}
    \caption {eRC eVTOL}
    \label{fig:erc}
\end{figure}

\subsection{Avilus Grille}
Avilus Grille \cite{o5} is an unmanned multicopter built for autonomous medical evacuation in hazardous or inaccessible areas. The fully electric aircraft with a MTOW of 695 kg, focuses on reliability and mission flexibility. Currently the aircraft undergo flight tests.
\begin{figure}[ht]
    \centering
    \includegraphics[width=0.49\textwidth]{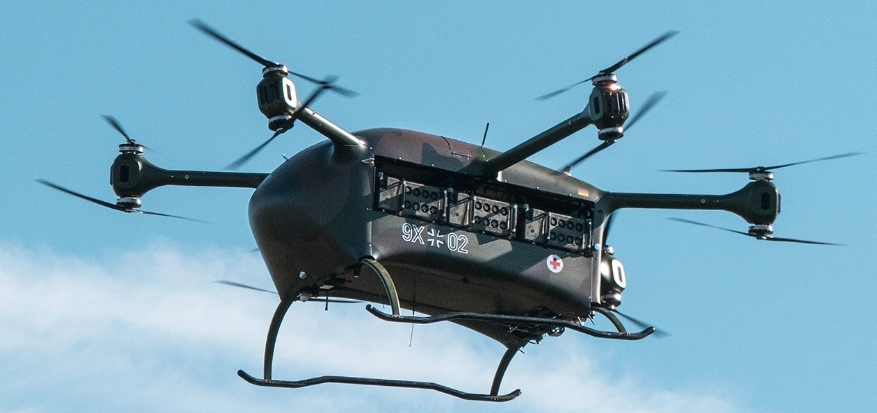}
    \caption {Grille Multicopter}
    \label{fig:grille}
\end{figure}

\subsection{AMI-PGS}
AMI-PGS is a small (total mass around 80 kg), experimental airship developed within the Air Mobility Initiative \cite{o6}. It is used to evaluate radar-based sensing technologies for applications such as ground surveillance and situational awareness. Its low weight and modularity make it ideal for technology evaluation within a built-up area.

\begin{figure}[ht]
    \centering
    \includegraphics[width=0.49\textwidth]{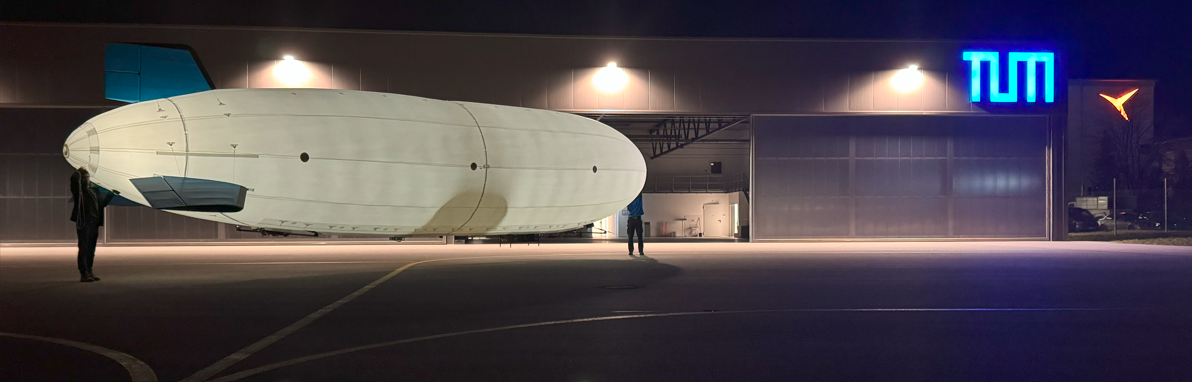}
    \caption {AMI-PGS Airship}
    \label{fig:amipgs}
\end{figure}

\subsection{HYFLUGS}
HYFLUGS is a modified ‘Breezer Sport’ fixed-wing light aircraft [23] with a MTOW of 600 kg modified to host a complex autopilot and flight control system in a mechanically redundant setup. Operating in the general aviation segment, it serves as a platform for testing hybrid flight control architectures and run-time assurance concepts. More details are available in \cite{b55}

\begin{figure}[ht]
    \centering
    \includegraphics[width=0.49\textwidth]{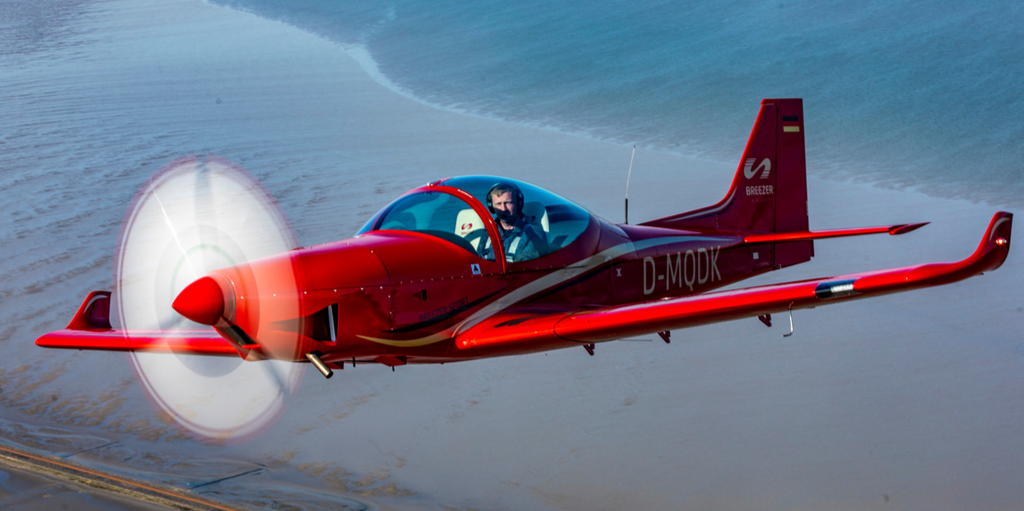}
    \caption {Breezer Sport Aircraft}
    \label{fig:hyflugs}
\end{figure}

\section{Conclusion and Future Work}

In this work, we have successfully developed, implemented, and refined a process that meets the goals outlined in Section~I: efficiency, transparency, and verifiability.

Efficiency was achieved by automating bulky, error-prone activities and emphasizing artifact reuse across multiple projects.  
Transparency was achieved by explicitly defining artifacts and establishing clear relationships between process steps.  
Verifiability was ensured through careful selection of tools and methods aligned with principles stipulated in standards such as ARP4754B and DO-178C.

An additional benefit of the process is the definition of clear, repeatable guidelines that can be readily adopted for new projects, improving scalability and consistency across developments.

While several important development activities lie outside the scope of this paper, it is worth noting their connection to the presented process:

\begin{itemize}
    \item Architecture Development and Function Allocation:
    Initial high-level architecture development, unit selection, and function allocation are performed before the start of the software development process.  
    However, once the detailed architecture design is refined through the population of ICD Database process, architecture artifacts can be automatically updated within the toolchain.

    \item Hardware-in-the-Loop (HIL) Based Verification:
    Verification of single units or complete systems using HIL platforms benefits from the same machine-readable ICD-based approach.  
    Input-Output (I/O) channel configuration for HIL simulations, as well as automated development of HIL rig EWIS (Electrical Wiring Interconnection System), is streamlined using XML-based ICD data, reducing manual effort and minimizing errors.

    \item Aircraft EWIS Development:  
    Aircraft wiring design is a natural extension of the toolchain, allowing automatic derivation of EWIS artifacts from the ICD database, further improving integration efficiency.
\end{itemize}

Future work will focus on the following activities:

\begin{itemize}
    \item Summarizing process details into formal plans according to ARP4754B and DO-178C, including Development, Validation, and Verification Plans.
    \item Advancing prototype tool developments towards formal TQL-5 qualification, enabling potential Design Assurance Level (DAL) C applications for automatically generated artifacts.
    \item Enhancing the connectivity between the process described and safety assurance processes, by enabling automated extraction of structured data from design artifacts, enabling their use as input to domain-specific tools for activities such as fault tree analysis (FTA), improving traceability and reducing manual effort in safety assessment processes.
\end{itemize}

\end{document}